\documentstyle[twoside,fleqn,epsfig,espcrc2]{article}
\newcommand{\ee}{\end{enumerate}}        
\newcommand{\bi}{\begin{itemize}}
\newcommand{\ei}{\end{itemize}}        
\newcommand{\eq}[1]{{\protect\frenchspacing eq.~(\ref{#1})}}

\newcommand{\beq}{\protect\begin{equation}}
\newcommand{\eeq}{\protect\end{equation}}        
\newcommand{\bqa}{\begin{eqnarray}}        
\newcommand{\eqa}{\end{eqnarray}}        
\newcommand{\be}{\begin{enumerate}}
\newcommand{\tauflip}{\tau_{\mbox{\tiny flip}}}

\newcommand{\vmuca}{V_{\mbox{\tiny MUCA}}}

\newcommand{\nn}{\nonumber}

\newcommand{\prl}[3]{\frenchspacing Phys. Rev. Lett. {\bf #1} {(#2)}
  {#3}}
\newcommand{\plb}[3]{ \frenchspacing Phys.  Lett.  {\bf #1B} {(#2)}
  {#3}}

\newcommand{\prd}[3]{ \frenchspacing Phys.  Rev.   {D\bf #1} {(#2)}
  {#3}}

\newcommand{\jcp}[3]{ \frenchspacing J.  Comp. Phys.   {\bf #1} {(#2)}
  {#3}}

\title{Multicanonical hybrid Monte Carlo for compact QED}
\author{
  G.~Arnold, K.~Schilling\\
  {\small NIC, c/o Research Center J\"{u}lich and DESY, Hamburg,
    D-52425
    J\"{u}lich, Germany}\\[8pt]
  {Th.~Lippert}\\
 {\small Department of Physics, University of
    Wuppertal, D-42097 Wuppertal, Germany}}
\begin{document}
\begin{abstract}
  We demonstrate that substantial progress can be achieved in the
  study of the phase structure of 4-dimensional compact QED by a joint
  use of hybrid Monte Carlo and multicanonical algorithms, through an
  efficient parallel implementation. This is borne out by the
  observation of considerable speedup of tunnelling between the
  metastable states, close to the phase transition, on the Wilson
  line. Our approach leads to a	
general parallelization scheme for the efficient stochastic sampling
of systems where (a part of) the Hamiltonian involves the total action
or energy in each update step.
\end{abstract}
\maketitle
\section{Introduction}
It is embarrassing that lattice simulations of compact QED still have
not succeeded to clarify the order of the phase transition near $\beta
= 1$, the existence of which was established in the classical paper of
Guth \cite{GUT78}. This is mainly due to the failure of standard
updating algorithms, like metropolis or heatbath to move the system at
sufficient rate between the observed metastable states near its phase
transition.  Due to supercritical slowing down (SCSD) the tunneling rates decrease exponentially in $L^3$ and
exclude the use of lattices large enough to make contact with the
thermodynamic limit by finite size scaling techniques (FSS)
\cite{FIS70}.\\

Torrie and Valleau \cite{TORRIE76} used arbitrary ``non-physical'' sampling
distributions in their method, termed ``umbrella sampling'', to improve the efficiency of stochastic sampling for
situations when dynamically nearly disconnected parts of phase space
occur by biassing the system to frequent the dynamically depleted,
connecting regions of configuration space.
Berg and Neuhaus \cite{BER91} applied the idea of ``umbrella
sampling''under the name ``multi-canonical algorithm'' (MUCA) to
the simulation of a variety of systems exhibiting first-order phase
transitions.  In this
procedure, the biassing weight of a configuration with action
$S$ is dynamically adjusted such as to achieve a
near-constant overall frequency distribution over a wide range of $S$
within a {\em single} simulation.

\section{Multicanonical Sampling (MUCA)}
{``Canonical'' Monte Carlo generates a sample of field
configurations,$\left\{ \phi\right\}$ according to the Boltzmann weight $P_{\mbox{\tiny can}}(\phi) \sim
e^{-\beta S(\phi)}$.
The canonical action density which in general
    exhibits a double peak structure  at a first-order phase
    transition, can be rewritten as
\begin{equation}
  N_{\mbox{\tiny can}}(S,\beta)=\rho(S)\, e^{-\beta S},
\label{CAD}
\end{equation}
with the spectral density $\rho(S)$ being independent of
the inverse temperature $\beta$.\\
The multicanonical approach aims at generating a flat action density
\begin{equation}
  N_{\mbox{\tiny MUCA}}(S,\beta_c)=\mbox{const., for }
  S_{\mbox{\tiny max1}}\le S\le S_{\mbox{\tiny max2}},
\end{equation}
in a range of $S$ that covers the double peaks located at
$S_{\mbox{\tiny max1}}$ and $S_{\mbox{\tiny max2}}$.
Therefore one modifies the sampling weights  introducing a multicanonical potential $V_{\mbox{\tiny MUCA}}(S)$,
\begin{equation}  
P_{\mbox{\tiny MUCA}}(S) \sim e^{-\beta_c S-V_{\mbox{\tiny
MUCA}}(S)}, 
\end{equation}
\bqa
V_{\mbox{\tiny MUCA}}(S)= \left\{
\small{\begin{array}{ll} 
\log{ N_{\mbox{\tiny can}}(S_{\mbox{\tiny max1}},\beta_c)} &
{\scriptstyle S<S_{\mbox{\tiny max1}}} \\
\log{N_{\mbox{\tiny can}}(S,\beta_c)} & {\scriptstyle S_{\mbox{\tiny max1}}\le S\le
S_{\mbox{\tiny max2}}}  \\
\log{N_{\mbox{\tiny can}}(S_{\mbox{\tiny max2}},\beta_c)} &
{\scriptstyle S>S_{\mbox{\tiny max2}}} 
\end{array}} \right. \nn
\label{vmucaEQ}
\eqa  
which is constant outside the relevant action range. Since
$V_{\mbox{\tiny MUCA}}(S)$ is unknown at the begin of the simulation,
it is instrumental for MUCA to bootstrap from good
guesstimates.  We shall do so by starting from an
observed histogram of the canonical action density,
$\hat{N}_{\mbox{\tiny can}}(S,\hat\beta_c)$, see \eq{CAD}, at the
supposed location of the phase transition, $\hat\beta_c$.  From the action
density, we compute $\hat V_{\mbox{\tiny MUCA}}(S)$.
The sampling then proceeds with the full MUCA weight,
\begin{eqnarray}
\hat P_{\mbox{\tiny MUCA}}(S) &\sim& e^{-\hat \beta_c S-\hat V_{\mbox{\tiny
MUCA}}(S)}\nn \\
&\sim& e^{-\big(\hat\beta_c+\hat\beta(S)\big) S -\hat\alpha(S)}.
\label{Pmuca}
\end{eqnarray}
In order to compute expectation values of observables, one
has to reweight the resulting action density to reconstitute the proper
canonical density $
P_{\mbox{\tiny can}}(S,\hat\beta_c) \sim \hat{P}_{\mbox{\tiny MUCA}}(S)\
e^{\hat{V}_{\mbox{\tiny MUCA}}(S)}$.
The computation of the multicanonical weights requires the knowledge of the global
and not just the local change in action for each single update step.
As a consequence, MUCA is not parallelizable for local update
algorithms. For remedy, we propose to utilize the HMC updating procedure.
 
\section{Hybrid Monte Carlo (HMC)}

In addition to the gauge fields $\phi_{\mu}(x)$ one introduces a set
of statistically independent canonical momenta $\pi_{\mu}(x)$, chosen
at random according to a Gaussian distribution
$\exp(-\frac{\pi^2}{2})$.  The action $S[\phi]$ is extended to a
guidance Hamiltonian 
\beq 
{\cal H}[\phi,\pi]=\frac{1}{2}\sum_{\mu,x}\pi_\mu^2(x)+\beta S[\phi].
\label{Hguidance}
\eeq 
Starting with a configuration $(\phi,\pi)$ at time $t=0$, the
system moves through phase space according to the equations of
motion
\beq 
\dot{\phi}_\mu = \frac{\partial {\cal H}}{\partial\pi_\mu} =
\pi_\mu, \quad \dot{\pi}_\mu = -\frac{\partial {\cal H}}{\partial \phi_\mu}
=-\frac{\partial} {\partial \phi_\mu}[\beta S],
\label{EOM}
\eeq
leading to a proposal configuration $(\phi^\prime,\pi^\prime)$ at time
$t=\tau$. Finally this proposal is accepted in a global Metropolis
step with probability
\begin{equation}
\label{Hacceptance}
P_{acc}=\min\left( 1,e^{-\Delta {\cal H}}\right),    
{\scriptstyle \quad \Delta {\cal H}={\cal H}[\phi^\prime,\pi^\prime]-{\cal H}[\phi,\pi]}.
\end{equation}

The equations of motion are integrated numerically with finite step
size $\Delta t$ along the trajectory from $t=0$ up to
$t=N_{\mbox{\tiny md}}\Delta t=\tau$.  Using the leap-frog scheme \cite{DUA87} as symplectic
integrator the discretized version of \eq{EOM} fulfills the {\em
detailed balance} condition  being time
reversible and measure
preserving. Each
integration step approximates the correct ${\cal H}$ with an error of
$O(\Delta t^3)$.

\section{Merging MUCA and HMC\\ for Compact QED (MHMC)}

We consider a multicanonical HMC for pure 4-dimensional $U(1)$ gauge theory
with standard Wilson action defined as
\bqa
S[\phi]=\sum_{x,\nu>\mu}\Big[1-\cos\big(\theta_{\mu\nu}(x)\big)\Big],\nn
\eqa
\[
\theta_{\mu\nu}(x)=\phi_\mu(x)+\phi_\nu(x+\hat{\mu})-\phi_\mu(x+\hat{\nu})-\phi_\nu(x).
\]
Eq.~\ref{Pmuca} suggests to consider an {\em effective action $\hat{S}$}
including the ``multicanonical potential'' $\vmuca$,
\beq
{\hat{S}=\hat{\beta}_c S+\hat\vmuca(S,\hat\beta_c)}=(\hat{\beta}_c+\hat{\beta}(S))S+\hat{\alpha}(S).
\label{Seff}
\eeq

We now define MHMC making use of the multicanonical potential as a driving
  term within molecular dynamics via the multicanonical Hamiltonian, 
$
{\cal H}=\frac{1}{2}\sum\pi^2+{\hat{S}},
$ 
inducing an additional drift term such that the resulting force is
given by
\beq%
{\scriptstyle
\dot{\pi}_\mu(x)=
\Big({\hat\beta_c}+{\hat\beta(S)}\Big) \sum_{\nu\not=
\mu}\Big[\sin\theta_{\mu\nu}(x-{\hat \nu})-\sin\theta_{\mu\nu}(x)\Big].
}
\eeq%

The MHMC is governed by the dynamics underlying the very two peak
structure: $\vmuca$ is repelling
the system out of the hot (cold) phase towards the cold (hot)
phase, thus increasing its mobility and enhancing flip-flop activity
\footnote{Note that MHMC requires the
computation of the global action (to adjust the correct multicanonical
weight, \eq{Pmuca}) at each integration step along the trajectory of
molecular dynamics to guarantee reversibility.}. 

\section{Tunneling Behaviour}

In order to quantify the efficiency of the MHMC, we introduce the {\em
average flip
time $\tauflip$} defined as the inverse number of the
sum of flips between the two phases multiplied by the total number of
trajectories. For reference, we additionally measured $\tauflip$ from
the Metropolis algorithm with reflection steps (MRS) which is
considered as a very effective local update algorithm for U(1) \cite{BUN94}.
{\small
\begin{center}
\begin{tabular}{|r|l|l|l|} 
\hline
L & $\beta$ & $\tauflip^{\mbox{\tiny MRS}}$ &
$\tauflip^{\mbox{\tiny MHMC}}$  \\ \hline\hline
6 & 1.001600 & 508(12) & 650(20)      \\ \hline
8 & 1.007370 & 1023(60) & 1173(50)    \\ \hline
10& 1.009300 & 2474(117)& 2006(84)    \\ \hline
12& 1.010143 & 5470(770) & 3260(440)   \\ \hline
14& 1.010668 & 16400(3300) & 5090(630) \\ \hline
16& 1.010753 & 44800(9700) & 6350(860) \\ \hline
\end{tabular}\\
$\tauflip$ for the various lattices of volume
$V=L^4$.
\end{center}}

With the results for $\tauflip$ on lattices up to $16^4$ we are in the
position to estimate the scaling behaviour of MHMC in comparison to
standard MRS updates. According to the expected exponential behaviour of
$\tauflip^{\mbox{\tiny MRS}}$ which, in the asymptotic regime $L \to
\infty$, should be given by $\tau_{\mbox{\tiny SCSD}}\sim
  \exp\left(2\sigma L^3 \right)$, we perform a $\chi^2$-fit
with the ansatz:
\begin{equation}
  \tauflip^{\mbox{\tiny MRS}}=a \ L^b \ e^{c L^3}.
\end{equation}
which yields $\chi_{\mbox{\tiny per
    d.o.f.}}^2=0.897$. As a result, we find a clear exponential SCSD
behaviour for the MRS algorihm. On the other hand, for the tunneling
    times of the MHMC, we expect a 
monomial dependence in $L$: 
\begin{equation}
\tauflip^{\mbox{\tiny MHMC}}=p \ L^q
\label{powerfit}
\end{equation}
The power law ansatz is well confirmed by the fit quality with
$\chi_{\mbox{\tiny per d.o.f.}}^2=0.795$.  
We also took the pessimistic ansatz and tried to detect a potentially
exponential increase of $\tauflip^{\mbox{\tiny MHMC}}$.  The
exponential fit gives $\chi_{\mbox{\tiny per d.o.f.}}^2=0.975$.\\
\centerline{\includegraphics[width=7.5cm]{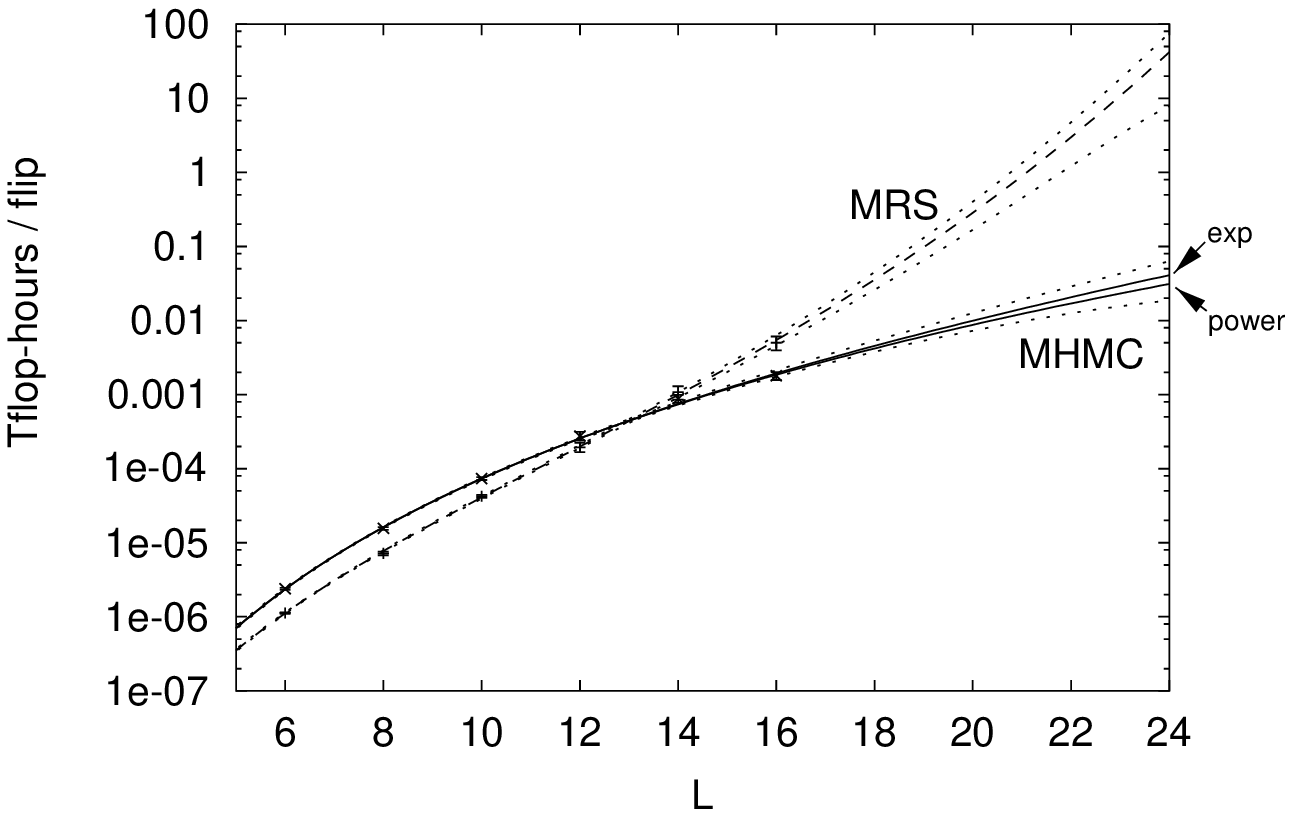}}
{\small
  Sustained CPU time in Tflop-hours to generate one flip based on
  tunneling times for MRS (exponential fit) and MHMC (lower curve is
power law, upper curve is exponential fit). The
  errors of the two exponential fits are depicted as dotted lines.
  The error of the power law fit is not visible on
  this scale.
  \label{algospeed}}\\
In order to compare the efficiency we have to take into account the
computational effort. The complexity of the
local Metropolis is given by $t_{\mbox{\tiny MRS}} \sim V$ whereas the optimized leapfrog
scheme in MHMC scales as $t_{\mbox{\tiny MHMC}} \sim V^{5/4}$ \cite{ARN98}.
As can be seen in the upper figure, the exponential contribution remains
suppressed in the extrapolation.  A potentially dominating exponential
behaviour for MHMC can only be detected in future MHMC simulations on
larger lattices.
\section{Conclusions}
We proposed to make use of the multicanonical (MUCA)
algorithm within the hybrid Monte Carlo (HMC) updating
scheme in order to boost the tunneling rates.  Since both
algorithms are inherently of global nature, their combination allows the parallelization of MUCA which could not be achieved
otherwise. We have demonstrated that the fully parallel MHMC algorithm is
capable to
overcome SCSD in compact QED in practical simulations, at least up to
lattices sizes $\approx 24^4$.
On a $24^4$ lattice we predict a gain factor of about 1000 for MHMC
over the local metropolis algorithm with additional reflection.
So far, we have encouraging
experiences on the $18^4$ lattice well confirming the extrapolations. 
The investigations presented form part of an ongoing study that aims at
a conclusive FSS analysis of compact QED on the Wilson line
\cite{ARN99}.


\begin{thebibliography}{9}
\bibitem{GUT78} A. H. Guth: \prd{21}{1980}{2291}.
%
\bibitem{FIS70} M. E. Fisher and M. N. Barber: \prl{28}{1971}{1516}.
%
\bibitem{TORRIE76} G. M. Torrie and J. P. Valleau: \jcp{23}{1977}{187}.
%
\bibitem{BER91} B. A. Berg and T. Neuhaus: \plb{267}{1991}{249}.
%
\bibitem{DUA87} S. Duane et al.: \plb{195}{1987}{216}.
%
\bibitem{ARN99} G. Arnold, Th. Lippert, Th. Neuhaus and K. Schilling: to appear.
\bibitem{BUN94} B. Bunk: proposal for U(1) update, unpublished,
  private communication.
%
\bibitem{ARN98} G. Arnold, Th. Lippert, and K. Schilling: \prd{59}{1999}{054509}

\end{thebibliography}
\end{document}